# Atomic investigations on the mechanical properties of CoCrNi medium-entropy alloy nanowires

Yang Yu[1]*

[1]*School of Physics and Optoelectronic Engineering, Nanjing University of Information Science and Technology, China*

*Electronic mail: yuyang@nuist.edu.cn

**Abstract:**
This work investigated the mechanical properties of CoCrNi medium-entropy alloy (MEA) nanowires under uniaxial tensile loading along three crystallographic orientations [100], [110] and [111]. The face-centered cubic (FCC) single-crystal CoCrNi nanowires exhibit pronounced anisotropy. Molecular dynamics simulations were employed to provide atomic-level insights into the plastic deformation mechanism. The surface effect causes dislocations to nucleate at the free surface and slip inward. The temperature effect was also considered. Under identical geometric configurations, crystal orientations, and temperatures, interatomic interactions govern key mechanical properties such as Young's modulus and yield strength. Simulation results reveal a strong linear correlation between the average atomic force and mechanical performance across eight representative FCC metallic nanowires (Al, Au, Ag, Cu, Ni, $Al_{19}Mg$ alloy, FeCoCrCuNi high-entropy alloy (HEA), and CoCrNi MEA). This study delivers novel insights into the mechanical behavior of metallic nanowires and offers guidance for the rational design of alloy nanowires.

1. Introduction:

Since the birth of CoCrFeMnNi in 2004[1, 2], multi-principle element alloys(MPEAs) have gained extensive attention in the past two decades and have achieved significant progress due to their super mechanical properties, such as ultrahigh strength and hardness, excellent high/cryogenic temperature performance, high fracture toughness and good ductility[3-9]. The key concept is not the number of components, but rather the near-equimolar composition of at least three major metallic elements that breaks through the design philosophy of conventional alloys (using one or two metallic elements as the base components and adding specific small amounts of other elements as minor modifications). This strategy brings alloy design to the central region of the phase diagram[10], and theoretically expands the possibilities for alloy design limitless. With the higher mixed entropy compared to the traditional alloys, the MEPAs inhibit the formation of intermetallic compounds and form solid solutions with a single crystal structure. That's why the MEPAs are given the name high-entropy alloy(HEA)[1] or medium-entropy alloys(MEA)[11].

Owing to the superior characteristics of nanomaterials in comparison to bulk materials, coupled with their extensive range of applications, the research emphasis on multi-component alloys has progressively transitioned into the nanoscale domain in recent years [12-19]. Due to the large surface area-to-volume ratio, unique electronic structure and superior mechanical properties, nanowires have a broad spectrum of applications related to the energy storage, catalysis, electronic communication, sensing, photonics, aerospace, transparent electrodes and nanoelectromechanical systems [19-24]. With the expansion of extreme application environments such as ultra-high or cryogenic temperature, high pressure, and corrosion-prone condition, there are special requirements for the materials used to fabricate the nanowires. The CoCrNi MEA is a good candidate for the nanowire material due to its merits of enhanced high-temperature stability, super low-temperature performance, superior strength-ductility combination and corrosion resistance.[3, 4, 25-29]

Mastering the mechanical properties of nanowires is of great significance for their applications, as they are often under stress during service. There are many factors that affect the characteristics of nanowires: elemental composition, cross-sectional geometry (shape and size) [30, 31], surface effect [32-35], crystallographic orientation [36-39], ambient temperature [40-42], applied strain rate [43-46] and so on. This work utilized molecular dynamics simulation to conduct a systematic investigation into the influents of surface effect, crystallographic orientation and ambient temperature on the deformation behavior and mechanical performance of CoCrNi single-crystal nanowire. The effects of different elemental compositions on the mechanical properties of metallic nanowires were also investigated.

## 2. Simulation details:

The LAMMPS (large-scale atomic/molecular massively parallel simulator)[47-49] software package was applied for all of the molecular dynamics (MD) simulations. The software OVITO[50] was used to visualize the simulation results as well as perform the common neighbour analysis (CNA) and dislocation analysis (DXA).

The EAM (embedded atom method) interaction potential developed by Farkas[51] was utilized to describe the interatomic forces within the CoCrNi alloy. This version of potential has been successfully employed to study the mechanical properties of the CoCrNi MEA [52-59]. For a comparison, a series of metallic nanowires was modelled. The EAM potentials used for these nanowires are Al[60], Au[61], Ag[62], Cu[63], Ni[64] , $Al_{19}Mg$ alloy[65] and FeCoCrCuNi HEA [51] respectively. For comparison of atomic forces, a set of alkali and alkaline‑earth metals, including Li[66], Na[66], K[66], Cs[66], Be[67], Mg[68] and Ca[69] were also modelled.

This study centers on FCC single-crystalline CoCrNi cylindrical nanowires with diameter D = 8 nm and length L = 30 nm. To investigate the dependence of mechanical properties and deformation mechanisms on axis orientation during the tensile process, the crystalline orientations along the NW axis were set as [100], [110], and [111], respectively. The corresponding simulation cells were constructed with the atom counts of 134,385, 133,875, and 134,701 in sequence. The schematic model is depicted in Fig.1. The Co, Cr and Ni elements were equal proportion and randomly distributed. The periodic boundary condition was applied along the cylinder axis direction to simulate an infinite length of the nanowires. The other two directions adopted free boundary conditions. The structure energy was minimized by the conjugate gradient method. Then the NWs were relaxed under NPT (isothermal-isobaric) ensemble for 100 ps at 300 K to reach equilibrium. After that the uniaxial tension load was applied along the NW axis at a strain rate $5 \times 10^8$/s. During the stretching, the NVT (canonical) ensemble was employed, with the temperature maintained at 300 K.

To study how temperature affects the deformation behavior of CoCrNi NW, the [100]-oriented

CoCrNi NWs were stretched at 77 K, 300 K and 800 K respectively. Concurrently, to explore the correlation between mechanical properties and chemical composition, simulations were carried out on [100]-oriented Al, Au, Ag, Cu, Ni, $Al_{19}Mg$ and FeCoCrCuNi NWs with geometric dimension and stretching condition identical to those of CoCrNi [100] NW at 300 K.

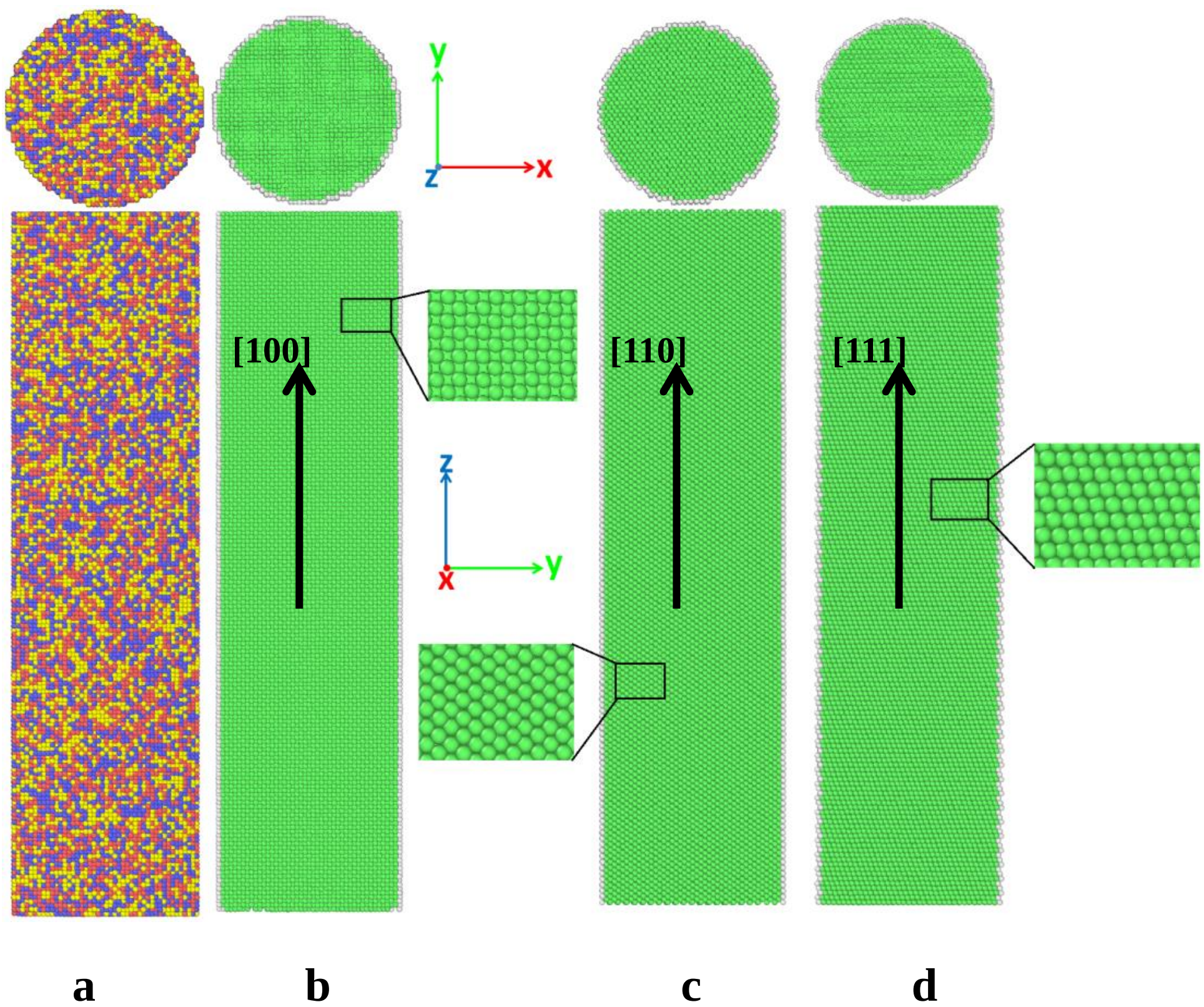

Fig. 1. Top and side views of the initial configuration of FCC single-crystalline CoCrNi nanowires. Circular insets at the top present the XY-plane cross-sections of the nanowires, while the side views are sliced along the Z-axis to expose their internal structures. (a) Spatial distribution of Co, Cr and Ni atoms. (Red:Co; Yellow:Cr; Blue:Ni). (b), (c) and (d) represent nanowires oriented along the Z-axis with [100] [110] and [111] crystallographic directions, respectively.

## 3. Result and discussion:

### 3.1. Surface effect in nanowires:

The mechanical properties of nanowires are highly sensitive to their surface structure [34, 70, 71]. Surface and subsurface atoms have fewer neighbors than interior lattice atoms, significantly altering their electron densities and energy states. This is evident from the spatial distribution of potential energy per atom across the cross-sections of CoCrNi and Ag nanowires, as illustrated in Figs. 2a and 2c. The [100]-oriented nanowires were fully relaxed to equilibrium at 300 K. For both nanowires, surface atoms exhibit higher potential energy than interior atoms, consistent with the per-atom potential energy histogram shown in Fig. 3.

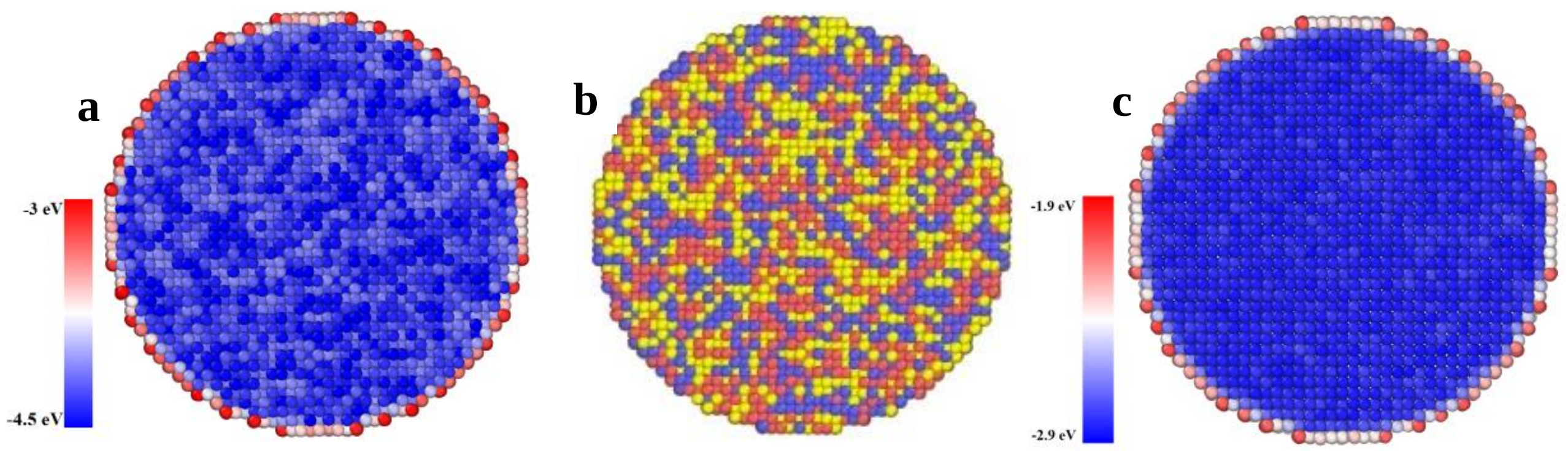

Fig. 2. Top view snapshots of the nanowires along the [100] crystallographic direction. (a) Potential energy nephogram of the CoCrNi nanowire. Atoms are colored according to their potential energy (unit: eV). (b) Atomic distribution of Co, Cr and Ni elements in the CoCrNi nanowire. (Red:Co; Yellow:Cr; Blue:Ni) (c) Potential energy map of the Ag nanowire.

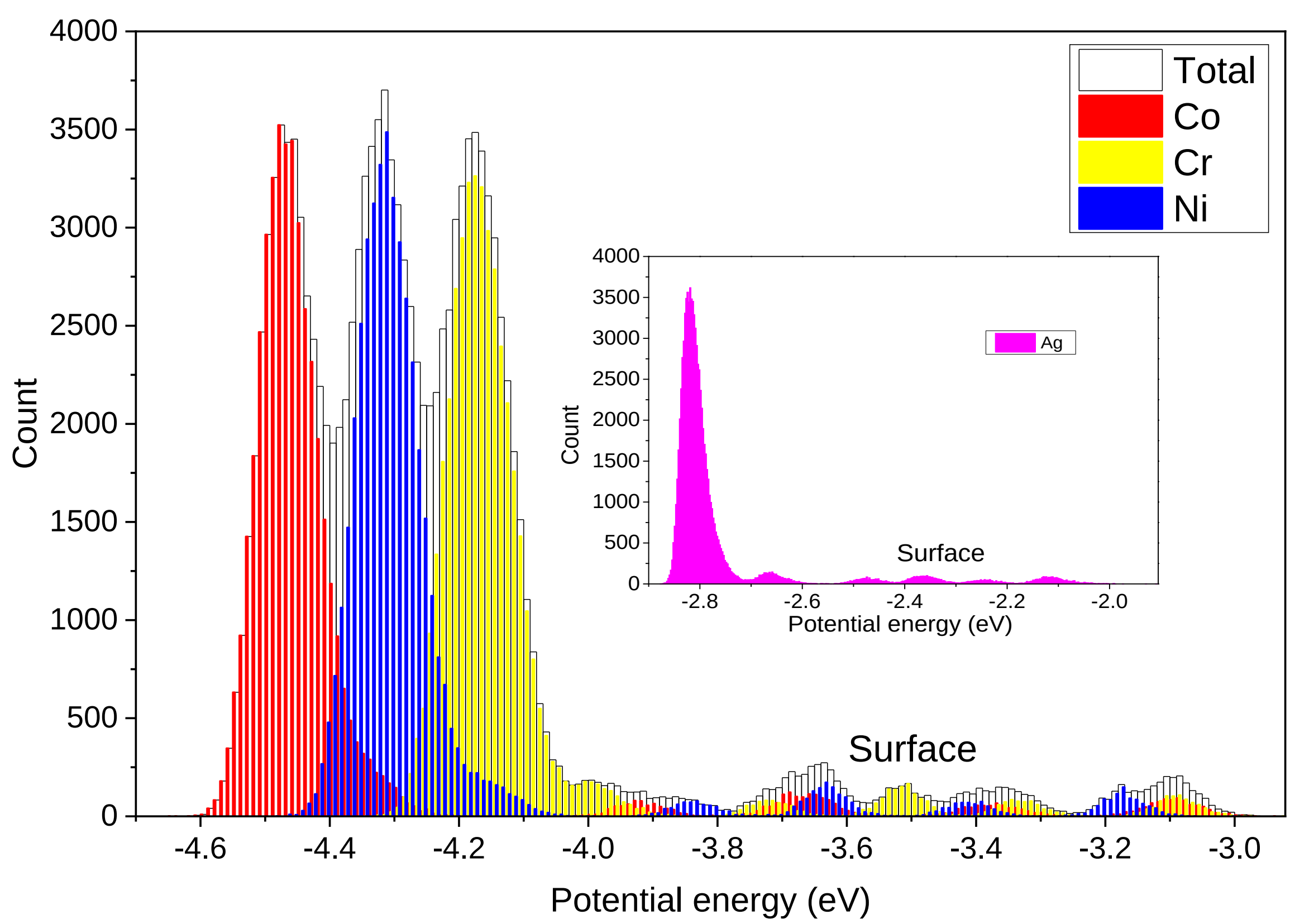

Fig.3. Per-atom potential energy histogram of the CoCrNi nanowire along the [100] crystallographic direction at 300 K. (Red: Co atoms; Yellow: Cr atoms; Blue: Ni atoms; White: total atoms). Inset: per-atom potential energy histogram of the Ag nanowire.

As shown in Figs. 2a and 2c, the CoCrNi alloy exhibits more significant fluctuations in potential energy than pure Ag. A comparison of the potential energy distribution and atomic distribution between Figs. 2a and 2b reveals that regions of higher potential energy generally correspond to Cr-rich domains, while regions of lower potential energy correspond to Co-rich domains. This correspondence is more clearly manifested in Fig. 3. Quantitative analysis of the potential energy histograms in Fig. 3 reveals that the CoCrNi alloy nanowire exhibits a significantly broader distribution than that of the Ag nanowire. The peaks corresponding to components Ni, Cr, and Co lie in close proximity, reflecting the similar energy values of these elements in their pure elemental states. The distinct chemical properties of the constituent alloying elements are the primary origin of the local inhomogeneity in the potential distribution of CoCrNi nanowire. Consequently, owing to local compositional fluctuations, the potential energy landscape of the CoCrNi nanowire is less homogeneous than that of the Ag nanowire. This rugged potential energy landscape in multi-principal element alloys (MPEAs), in contrast to the relatively smooth landscape of single-element metals, accounts for their enhanced mechanical strength. Specifically, the large spatial fluctuations in energy effectively impede dislocation motion, thereby contributing to the superior strength observed in these alloy systems. [72-74]

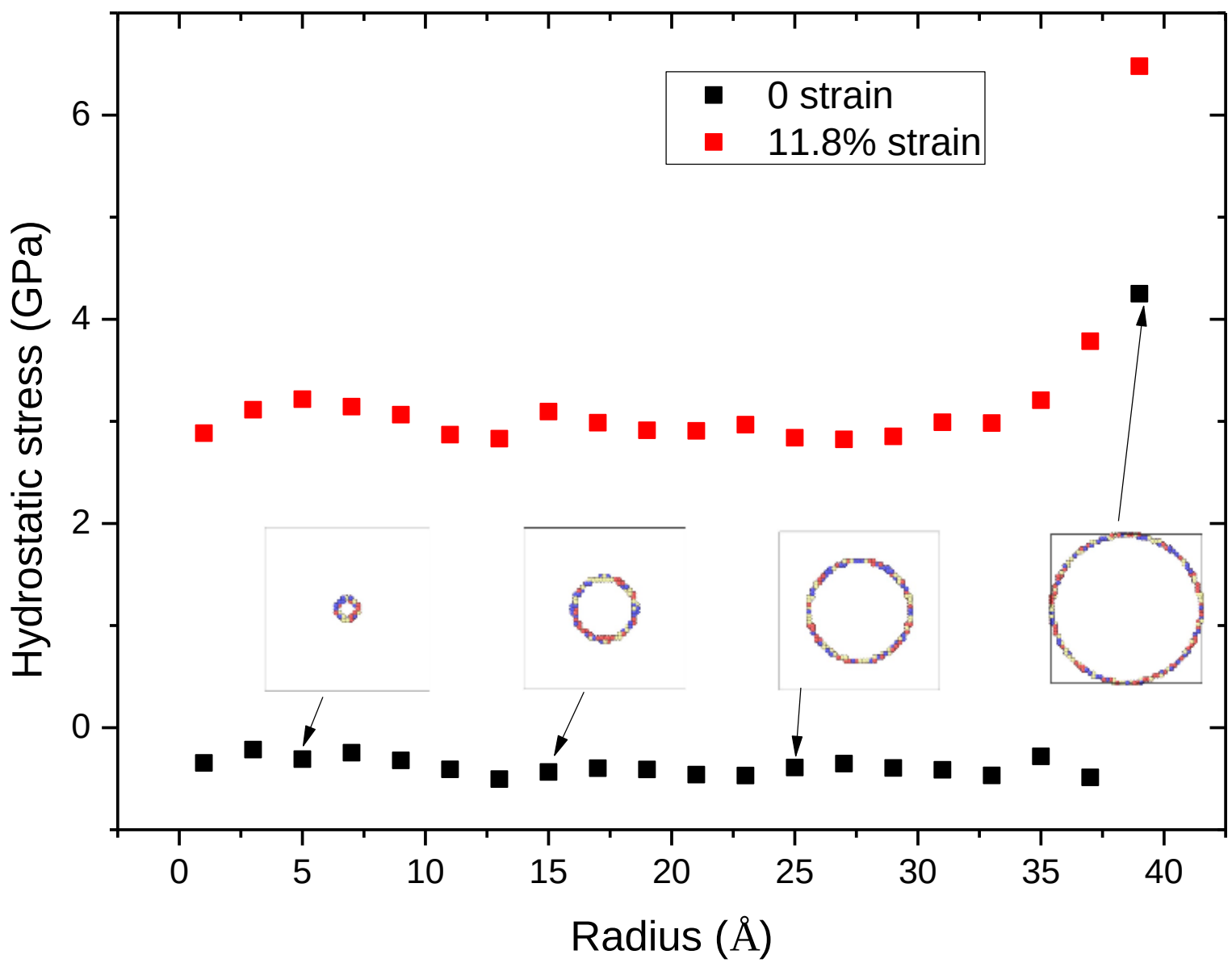

Fig. 4. Average per-atom hydrostatic stress of each layer along the radial direction in the [100]-oriented CoCrNi nanowire at 300 K (Black: unstrained state; Red: 11.8% strain, corresponding to the onset of deformation). Inset: Top view of atoms within several distinct layers adopted for stress calculations.

Besides energy disparities, surface atoms have lower coordination numbers, which lead to stress and strain states different from those of interior atoms. To compensate for the coordination deficiency, surface atoms tend to relax inward, giving rise to the free relaxation[75]. This behavior increases local stress and contracts bond lengths for surface atoms relative to those in the interior, resulting in pronounced stress concentration at the free surface [34, 75]. To quantitatively elucidate the stress variation from the nanowire core to its free surface, Fig. 4 displays the average per-atom hydrostatic stress of each layer along the radial direction of the CoCrNi nanowire. The zero-strain state corresponds to the equilibrium configuration without external loading, while the 11.8% strain state corresponds to the tensile configuration at the onset of dislocation nucleation. For both states, the stress in the outermost layer is substantially higher than that in the inner layers. Notably, for the zero-strain state, the inner region exhibits compressive stress, whereas the free surface exhibits tensile stress, consistent with the general characteristics of surface-induced residual stress fields in nanoscale materials[34].

Surface atoms possess higher potential energy, making them energetically less stable and requiring less energy to rupture their atomic bonds. Under elevated stress, surface atoms separate more readily than bulk atoms. The synergy of high local stress and potential energy provides preferential sites for dislocation nucleation. Consequently, dislocations tend to nucleate at the free surface rather than in the defect-free interior of a perfect crystal[35]. Simulation results directly confirmed this conclusion. As shown in the Supporting Information videos depicting the plastic deformation process from the yield point to the lower yield point, dislocations consistently nucleate at the free surface and glide inward in the CoCrNi nanowires for all three crystallographic orientations ([100], [110], and [111]).

### 3.2. Mechanical response of CoCrNi nanowires under tensile loading along different crystallographic orientations

Fig. 5 displays the stress-strain response of CoCrNi nanowires with three crystallographic orientations under uniaxial tension at 300 K. All curves rise quasi–linearly in the elastic regime, and then reach a maximum value corresponding to the yield strength, followed by an abrupt drop to lower yield points. A sawtooth pattern then appears in the plastic region. The CoCrNi nanowires show strong anisotropic dependence on tensile orientation. The Young’s modulus, yield stress and yield strain are listed in Table 1. A clear directional dependence is observed for the elastic modulus: $E_{[111]} > E_{[110]} > E_{[100]}$. This relationship is commonly found in FCC metal nanowires such as Ag[36], Al[37], Cu[38, 76, 77], Ni[78] and Au[79]. In FCC crystals, the (111) plane is close-packed plane. The higher the atomic density on a given plane, the greater the total interfacial force across the layer interface. As a result, a larger external force must be applied to overcome this interfacial attraction and achieve layer separation. When loading is applied along the <111> direction, it acts to separate the most densely packed atomic plane. The <110> direction is the close-packed direction in an FCC crystal, and the interlayer distance along this direction is the shortest. Consequently, a force applied along this direction separates more layers per unit length. Higher atomic density within the plane corresponds to enhancement at the two-dimensional level, while more layers along the tensile axis corresponds to enhancement at the one-dimensional level. Consequently, the <111> direction is more resistant to tensile separation than the <110> direction. The discussion above provides a qualitative explanation for the anisotropy of Young’s modulus in FCC crystals. A more rigorous quantitative treatment should take into account the elastic constants[80].

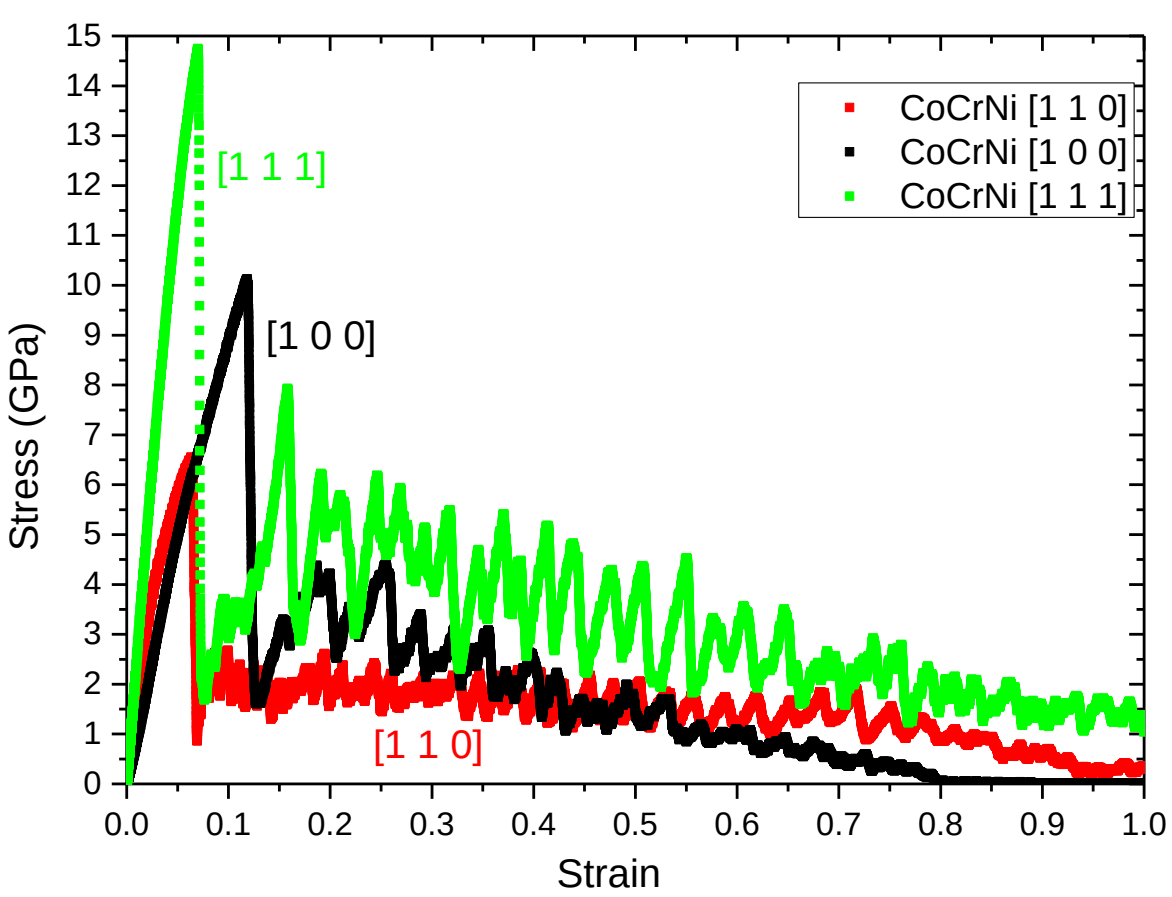

Fig. 5. Uniaxial tensile stress-strain curves of CoCrNi nanowires along different crystallographic orientations at 300 K.

Table 1. Young's modulus, yield stress and corresponding yield strain of [100], [110] and [111] oriented CoCrNi nanowires under tensile loading at 300 K.

| Nanowire's orientation | Young's modulus (GPa) | Yield stress (GPa) | Yield strain |
|---|---|---|---|
| [100] | 86.82 | 10.12 | 0.119 |
| [110] | 100.34 | 6.53 | 0.064 |
| [111] | 216.53 | 14.75 | 0.070 |

Table 2. Schmid factors and resolved normal stress factor of partial dislocations in FCC crystals for wires of various orientations under tension. Predicted mechanisms based on Schmid factors are also listed [39].

| Nanowire's orientation | Leading partial | Trailing partial | cosφ | Pridiction |
|---|---|---|---|---|
| [100] | 0.24 | 0.47 | 0.58 | f-slip |
| [110] | 0.47 | 0.24 | 0.82 | twin/p-slip |
| [111] | 0.31 | 0.16 | 0.33 | twin/p-slip |

The yield strength of CoCrNi nanowires exhibits strong orientation dependence $\sigma_{y[111]} > \sigma_{y[100]} > \sigma_{y[110]}$, as shown in Fig. 5, which is consistent with earlier reported results for Au[79] and Cu[77]. The yield strength is strongly correlated with the deformation mechanism . Fig.6 (a1), (b1) and (c1) illustrate the initial dislocation nucleation in nanowires of all three orientations. In each case, plastic deformation initiates via 1/6<211> Shockley partial dislocations that nucleate at the free surface and propagate inward. The corresponding Schmid factors[81, 82], calculated from the work of Weinberger et al.[39], are summarized in Table 2. According to the data, the leading partial dislocation in the [110]-oriented nanowire has the highest Schmid factor, indicating the lowest resolved shear stress required for dislocation activation. This result agrees well with this work and appears to be a general behavior of FCC single-crystal nanomaterials[39]. However, the results for the other two nanowires with [100] and [111] orientations are inconsistent with the Schmid factor predictions. This discrepancy can be attributed to the effect of normal stress on the (111) slip plane, which is not considered in the Schmid factor calculations. The dislocation activation force is analogous to frictional force: when an external force is applied at an angle to the slip plane, the in-plane force component must overcome the frictional force to initiate motion, whereas the frictional force itself scales with the normal pressure acting on the plane. According to the work of Li and Mishin [83], the stress component normal to (111) plane can strongly influence the stacking fault (SF) energy, and normal tension significantly reduces both the stable and unstable SF energy. The unstable SF energy serves as an energy barrier for the nucleation of a leading partial dislocation [39, 84], as the unstable SF energy decreases, the critical shear stress required for partial dislocation nucleation is correspondingly reduced. In other words, a greater normal tension leads to a lower critical shear stress.

Further observation of Fig. 5 reveals that the stress-strain curves of CoCrNi nanowires with different crystallographic orientations exhibit different mechanical behaviors in the plastic deformation stage. To elucidate the orientation-dependent deformation mechanisms, the atomic configurations of CoCrNi nanowires at the lower yield points are presented in Fig.6($a_2$), ($b_2$) and ($c_2$). A more detailed evolution process of atomic structures from the initial yield point to the lower yield points during plastic deformation is provided in the Supporting Information videos. For the [100]-oriented nanowire, leading partial dislocations first nucleate, followed by trailing partial dislocations, thereby forming full dislocations. The plastic deformation mechanism of [100]-oriented nanowire is dominated by both full and partial dislocations, with multiple slip systems being activated. The propagation of full dislocations generates steps on the free surface. Fig.6($a_3$) shows significant distortion induced by full dislocations. According to Schmid factor analysis, [110]-oriented nanowire under tension are expected to deform via twinning [39, 85]. Consistent with this theoretical prediction, the [110]-oriented CoCrNi nanowire in the present work exhibits a predominant twinning deformation behavior, as verified by the atomic configuration in Fig.6($b_2$). Additionally, partial-reorientation can be observed in Fig.6($b_3$). For the [111]-oriented nanowire, deformation is characterized by partial dislocations, slip, and stacking faults, with limited twining. Multiple slip systems are activated, as shown in Fig. 6($c_2$). Upon simultaneous activation of multiple slip planes, dislocations intersect and interact with one another during plastic deformation. The partial dislocations make small steps on the free surface as shown in Fig.6($c_3$). These deformation mechanisms are generally consistent with the Schmid factor predictions proposed by Weinberger et al.[39].

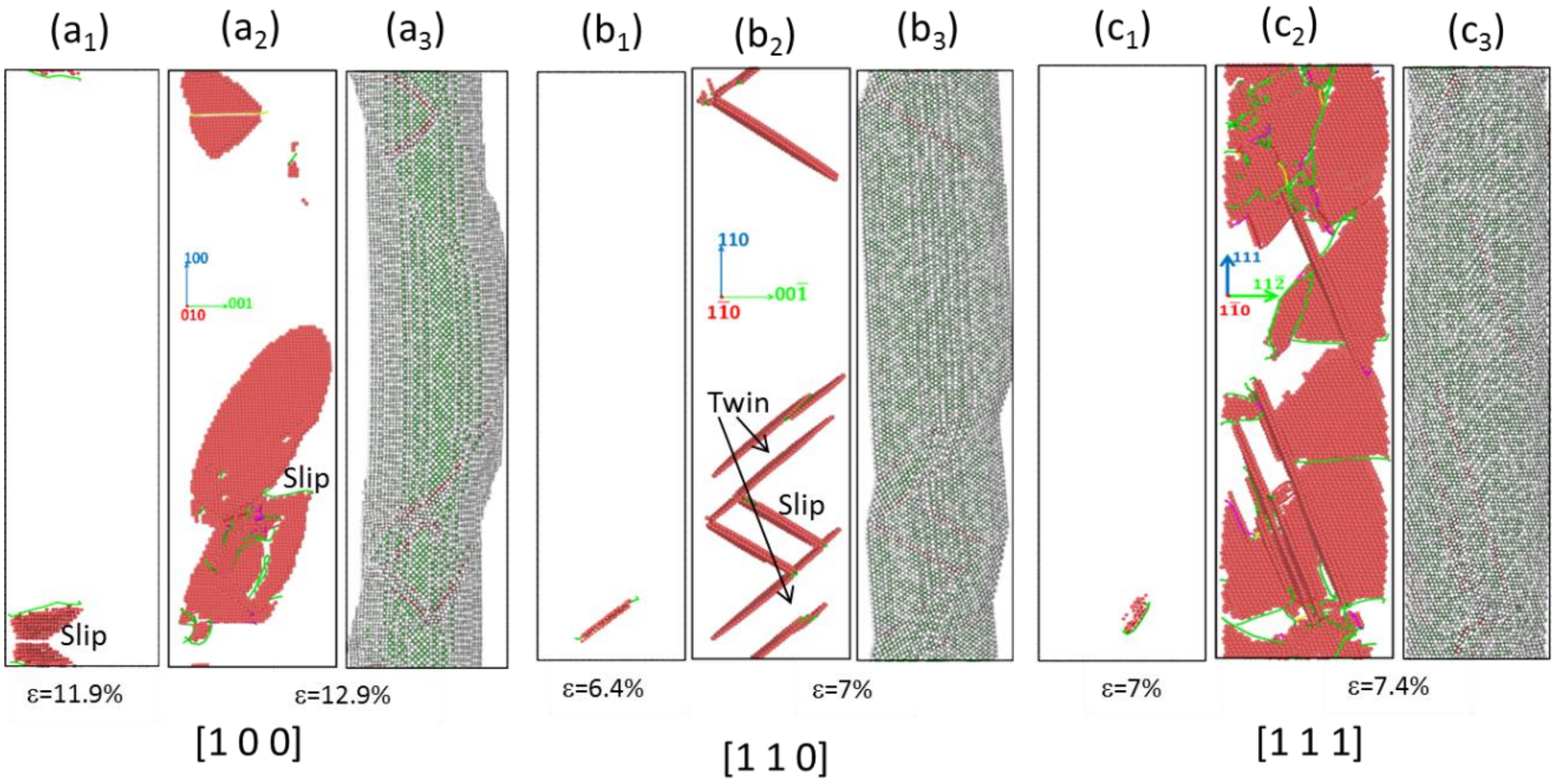

Fig. 6. Atomic configurations of CoCrNi nanowires with different crystallographic orientations at the yield point ($a_1$, $b_1$ and $c_1$) and the lower yield points ($a_2$,$a_3$, $b_2$,$b_3$ and $c_2$,$c_3$). Dislocation Analysis (DXA) was employed to analyse defect structures. (For visual clarity, only HCP-structured atoms are displayed, and Shockley partial dislocations are marked by green lines). Common Neighbor Analysis (CNA) was used to give the full image of nanowire deformation at the lower yield point ($a_3$,$b_3$ and $c_3$).

### 3.3. Mechanical response of CoCrNi nanowire at different temperatures

Temperature is a critical factor influencing the mechanical properties and deformation behavior of metallic materials. In this study, [100]-oriented CoCrNi nanowires were investigated at three representative temperatures: cryogenic temperature (77 K), room temperature (300 K) and a high temperature below melting (800 K). The corresponding stress-strain curves are shown in Fig. 7, which exhibit a regular pattern of variation. Specifically, yield strength increase gradually as temperature decreases, while Young's modulus shows a slight decrease with rising temperature. This behavior is generally observed in nanowires [77, 86, 87]. The increase in yield strength with decreasing temperature aligns with classical nucleation theory[88], which attributes this to the thermally activated nature of the dislocation nucleation process.

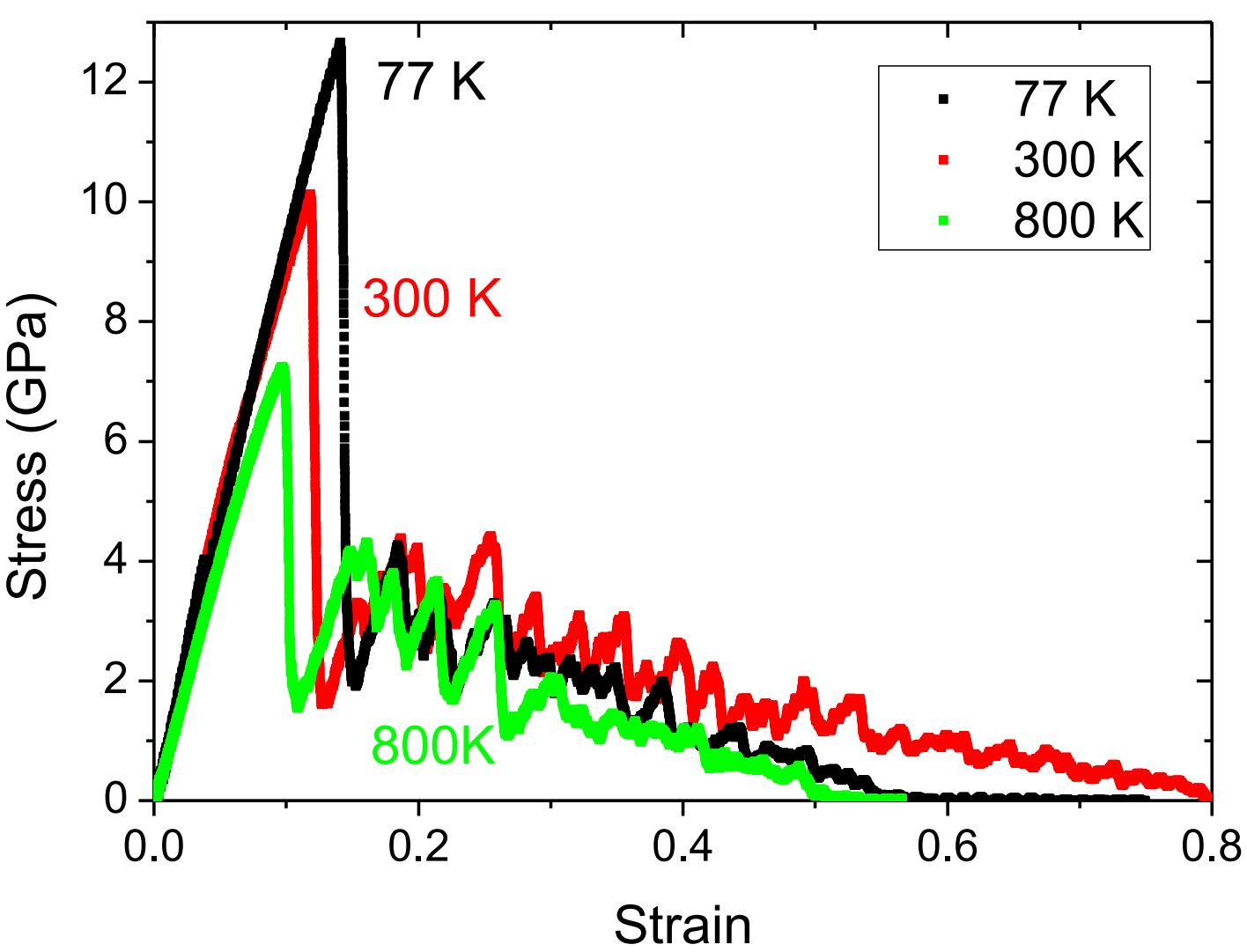

Fig. 7. Stress-strain curves of [100]-oriented CoCrNi nanowire under uniaxial tension at different temperatures.

### 3.4. Comparison of the mechanical response of CoCrNi nanowires with other metals

Various physical characteristics exhibited by matter at the macroscopic scale are determined by the inter-particle interaction mechanisms at the microscopic level. The macroscopic mechanical properties depend on the interatomic bonding ability at the microscopic level. The interatomic force changes with the interatomic separation[89]. Fig.8 is a schematic picture of the pairwise force F versus $r/r_0$. There is an equilibrium spacing $r_0$ that with this separation, net force equals 0. Since the equilibrium interatomic distance varies across various materials, the normalized ratio $r/r_0$ is adopted as the independent variable instead of the absolute distance for scaling purpose. The mechanical stiffness (or modulus of elasticity) of a material is dependent on the shape of its force–

versus–interatomic separation curve (slope at $r/r_0$=1). In the vicinity of the equilibrium spacing, the interatomic force varies approximately linearly with atomic separation, which is the physical origin of linear elastic behavior described by Hooke's law. When temperature rises, atomic motion becomes more vigorous. This leads to volume expansion and an increase in the average interatomic distance. Therefore, the $r/r_0$ on the horizontal axis can also be mapped to temperature. In the vicinity of $r/r_0$, the curve is approximately linear. So the pairwise atomic force at the equilibrium position at 300 K reflects the local slope associated with Young's modulus. As the first derivative of potential energy, this force serves as one indicator of the strength of interatomic interactions. Nevertheless, due to varying thermal expansion coefficients across metals, the force at 300 K provides only a qualitative, rather than quantitatively rigorous measure of interatomic bonding strength. Since the linear thermal expansion coefficient of metals is on the order of $10^{-6}$ $K^{-1}$, then $r/r_0 < 0.01$ at 300 K. When the differences in force are sufficiently large, they can override the variations in linear thermal expansion rates among different metals at 300 K.

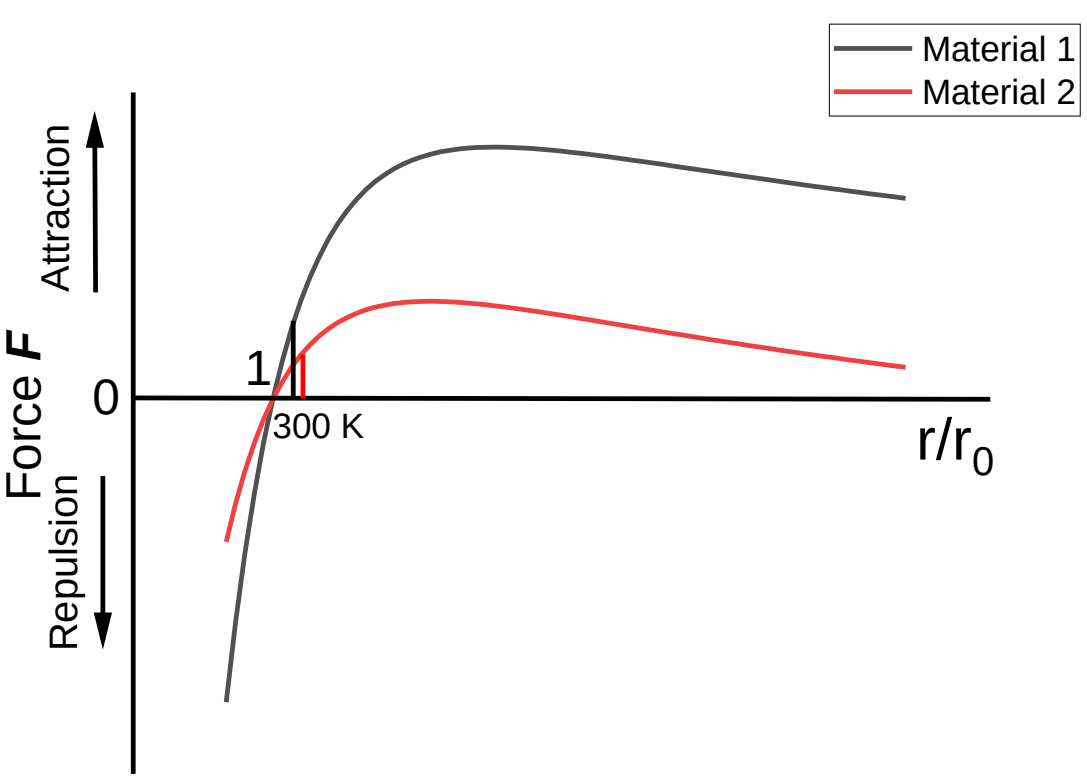

Fig.8 Schematic diagram of the pairwise interatomic force F as a function of the normalized interatomic distance $r/r_0$, where $r_0$ is the equilibrium interatomic spacing. In the vicinity of $r/r_0$=1, force varies linearly with displacement, corresponding to the linear elastic regime.

In a perfect crystal, because of the symmetry, the net force on an atom should be zero, since the force exerted by one neighbor is offset by another neighbor in the symmetrical position. But when in the actual crystal at room temperature, the thermal motion causes atoms to deviate from their equilibrium positions and gain excess force. This excess force should correlate with the interatomic force. This can be seen from Fig. 9. For both Ag and CoCrNi, the average force per atom increases with temperature, and the CoCrNi with stranger bonding strength shows higher force than Ag. These changes are similar to those shown in Fig. 8. Here, the mean value of the net force acting on each atom was calculated as $f = \frac{\Sigma\sqrt{f_x^2+f_y^2+f_z^2}}{N}$, where $f_x$,$f_y$ and $f_z$ components for each atoms were extracted from the LAMMPS simulation. The bulk structure was used in the simulation to eliminate complex factors of the nanostructure, such as free surface effects. Therefore, the average atomic force derived from the simulation of bulk structure at 300 K is adopted as a measure of bonding strength, in place of the pairwise atomic force.

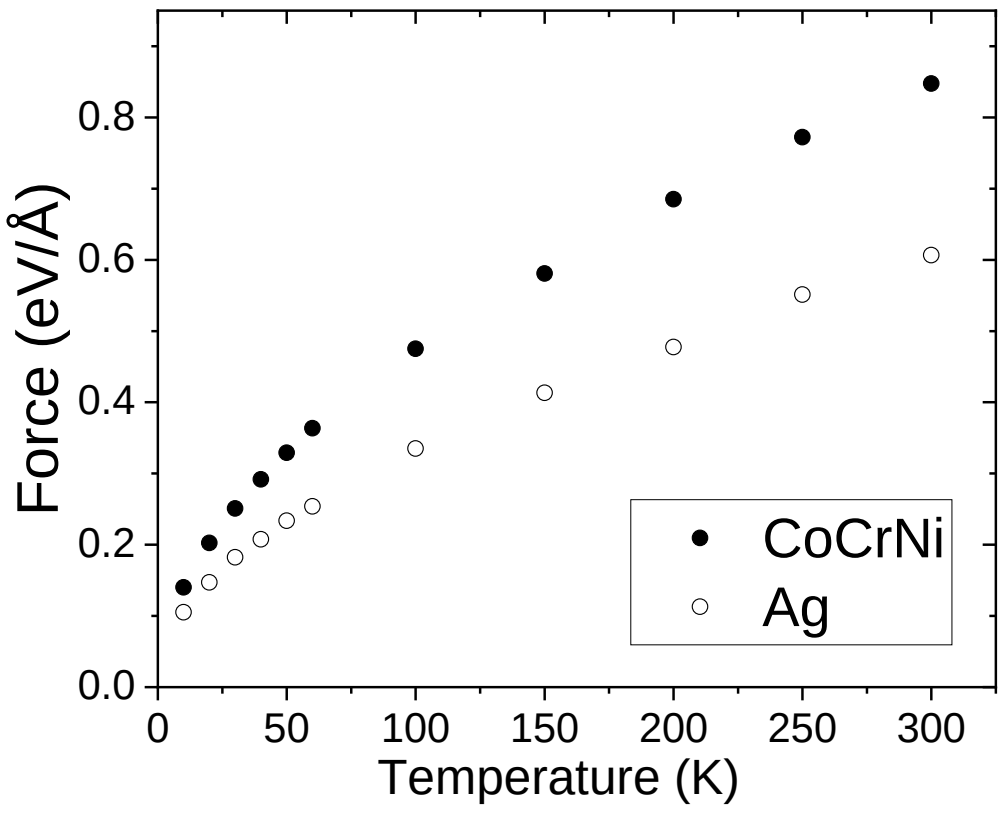

Fig. 9 Average force per atom for CoCrNi and Ag FCC bulk at different temperatures (10 K to 300 K) calculated using LAMMPS.

Fig. 10 and Fig.11show the correlation between macroscopic mechanical quantities and the microscopic atomic force. The stress-strain curves of the metallic nanowires are provided in the Supporting Information. As shown in Fig.10, there is a clear linear correlation between Young's modulus and the average atomic force, which is consistent with the previous analysis. In perfect crystals, dislocations are absent prior to the yield point. Young's modulus originates from the collective atomic interactions within the lattice, whereas plastic deformation following yielding is highly dependent on deformation mechanisms. In Fig. 11, the data point for Al deviates obviously from the linear trend. As a typical metal with high stacking-fault energy[90], Al is therefore expected to exhibit distinct behaviour compared with other metals of low stacking-fault energy [3, 90]. Nevertheless, it is unexpected that strength exhibits obvious linear dependence on the average force per atom for the other metallic nanowires in this work. Since the plastic behavior and yield strength of metals are governed by the resistance to dislocation slip, rather than directly by the strength of metallic bonds. One plausible interpretation is that the resistance to dislocation slip is correlated with atomic bonding strength.

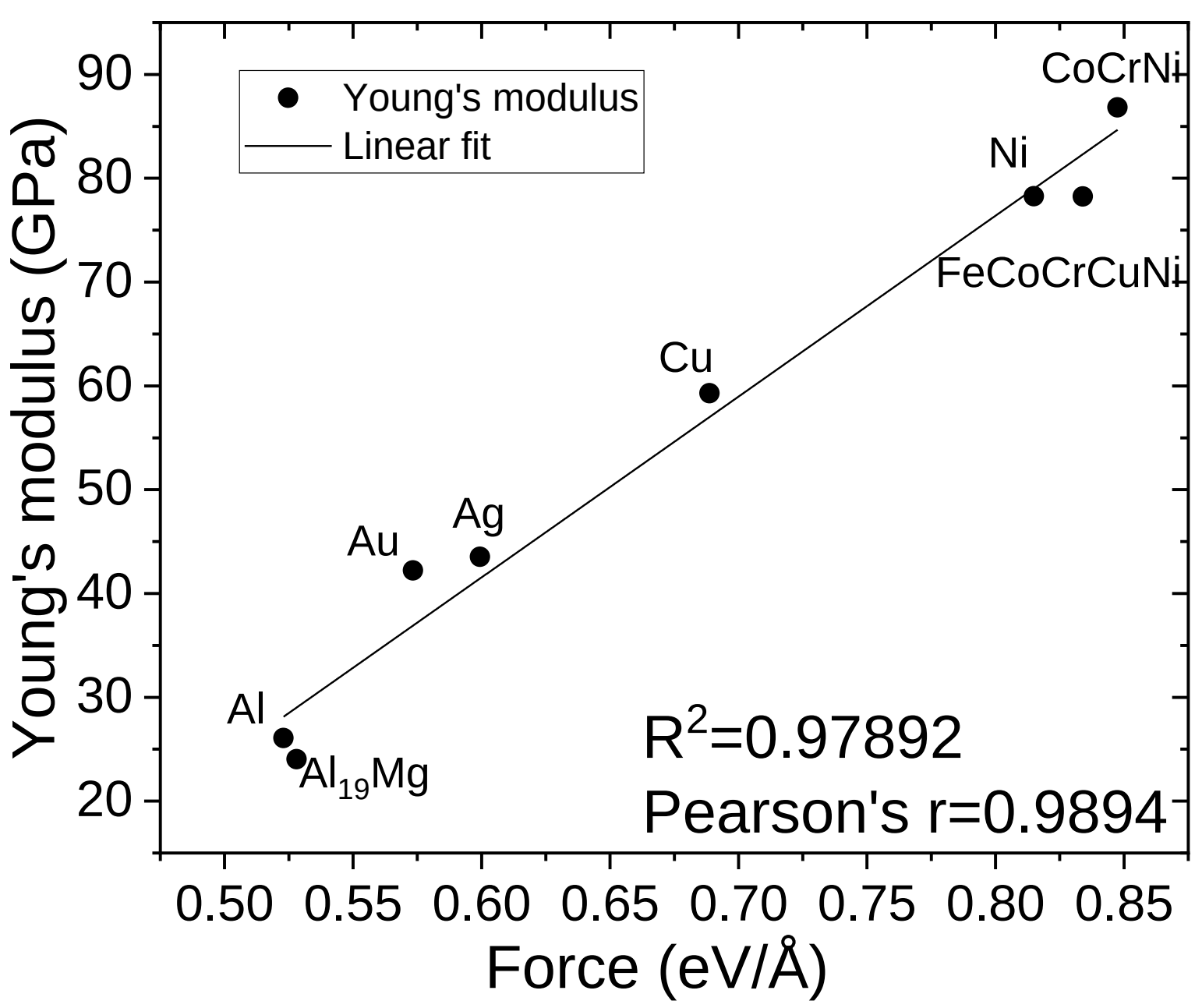

Fig.10. Young's modulus of various metallic nanowires at 300 K as a function of the average force per atom in the bulk calculated from LAMMPS.

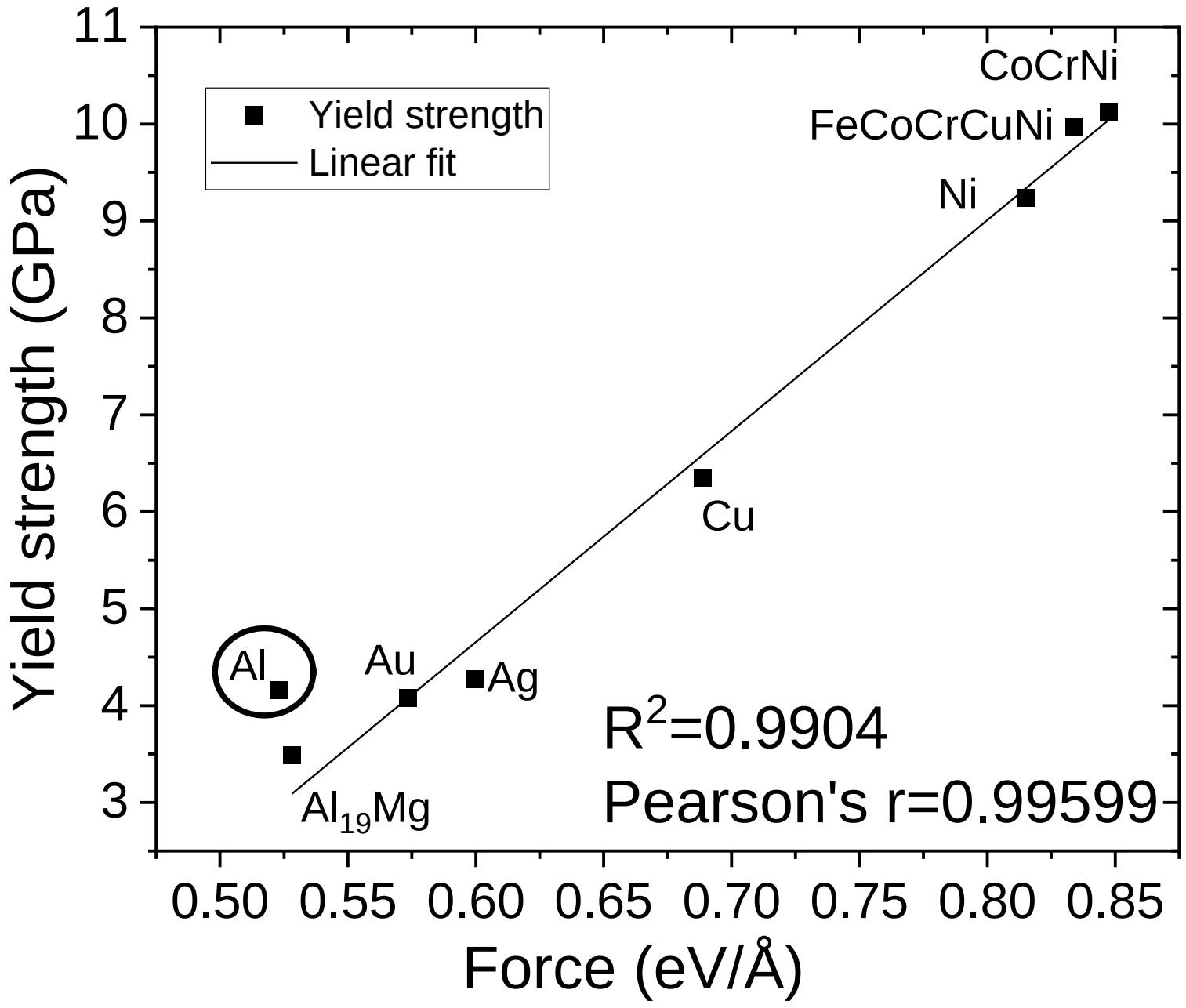

Fig.11. Yield strength of various metallic nanowires at 300 K as a function of the average force per atom in the bulk calculated from LAMMPS. Due to the large deviation of Al, this data point was excluded from the linear fitting.

From Fig. 10 and Fig. 11, the simulation results confirm that the strength of metallic bond serves as the determining factor for Young's modulus, and constitutes one of the key factors for yield strength. This macroscopic mechanism is correlated with microscopic atomic bonding and is also applicable to alloys in the nanoscale regime. Such force dependence can act as an evaluation criterion for selecting materials to meet specific design objectives.

There are many factors that influence the bonding strength of metallic materials. It is well known that the valence (Z) and atomic radius ($r_a$) of an element govern its metallic bonding strength. Fig.12 shows the relationship between the average force per atom and the electrostatic parameter $Z/r^2_a$. The values of Z and $r_a$ in this work are listed in Tabel 3[89]. For alloys, this parameter can be calculated via the rule of mixtures. Alkali and alkaline−earth metals Li, Na, K, Cs, Be, Mg, and Ca follow the linear correlation, while other metals deviate markedly from this trend. Several factors account for this discrepancy. First, valence electrons are not equivalent to free electrons in metallic solids for the transition metals, so the electrostatic parameter $Z/r^2_a$ is not correct here. Second, transition metals possess covalent contributions to metallic bonding and exhibit greater electronegativity compared with alkali and alkaline−earth metals. Accordingly, electronic structure plays a key role in determining metallic bonding strength. Metallic elements in the s region exhibit different mechanical responses to electronic structure compared with those in the p region and d region. For the transition metals, only a weak correlation is observed between the average force per atom and the electrostatic parameter $Z/r^2_a$. For elements with similar electronic configurations, valence and ionic radius are minor factors that influence the strength of metallic bond. To evaluate the electrostatic effects, the force of the $Co_3Cr_4Ni_3$ alloy was computed, with the material retaining an FCC crystal structure as a prerequisite. In this alloy, Cr exhibits the highest valence. Nevertheless, the atomic force of $Co_3Cr_4Ni_3$ is lower than that of CoCrNi. A plausible interpretation is that interatomic bonds between dissimilar elements are stronger than those between identical atoms. Increasing the concentration of one atomic species reduces the quantity of dissimilar element bonds, which in turn lowers the average force per atom within the material. Consequently, a simple electrostatic model relying solely on cation charge and radius cannot quantitatively explain the discrepancies. It is necessary to analyze the orbital bonding contributions via first-principles electronic structure calculations.

Table 3. Valence and atomic radii of the elements used in this work[89]

| Element | Al | Cr | Co | Cu | Au | Fe | Ni | Ag |
|---|---|---|---|---|---|---|---|---|
| Most common valence | 3 | 3 | 2 | 1 | 1 | 2 | 2 | 1 |
| Atomic Radius | 0.143 | 0.125 | 0.125 | 0.128 | 0.144 | 0.124 | 0.125 | 0.144 |
| Element | Li | Na | K | Cs | Be | Mg | Ca | |
| Most common valence | 1 | 1 | 1 | 1 | 2 | 2 | 2 | |
| Atomic Radius | 0.152 | 0.186 | 0.231 | 0.265 | 0.114 | 0.16 | 0.197 | |

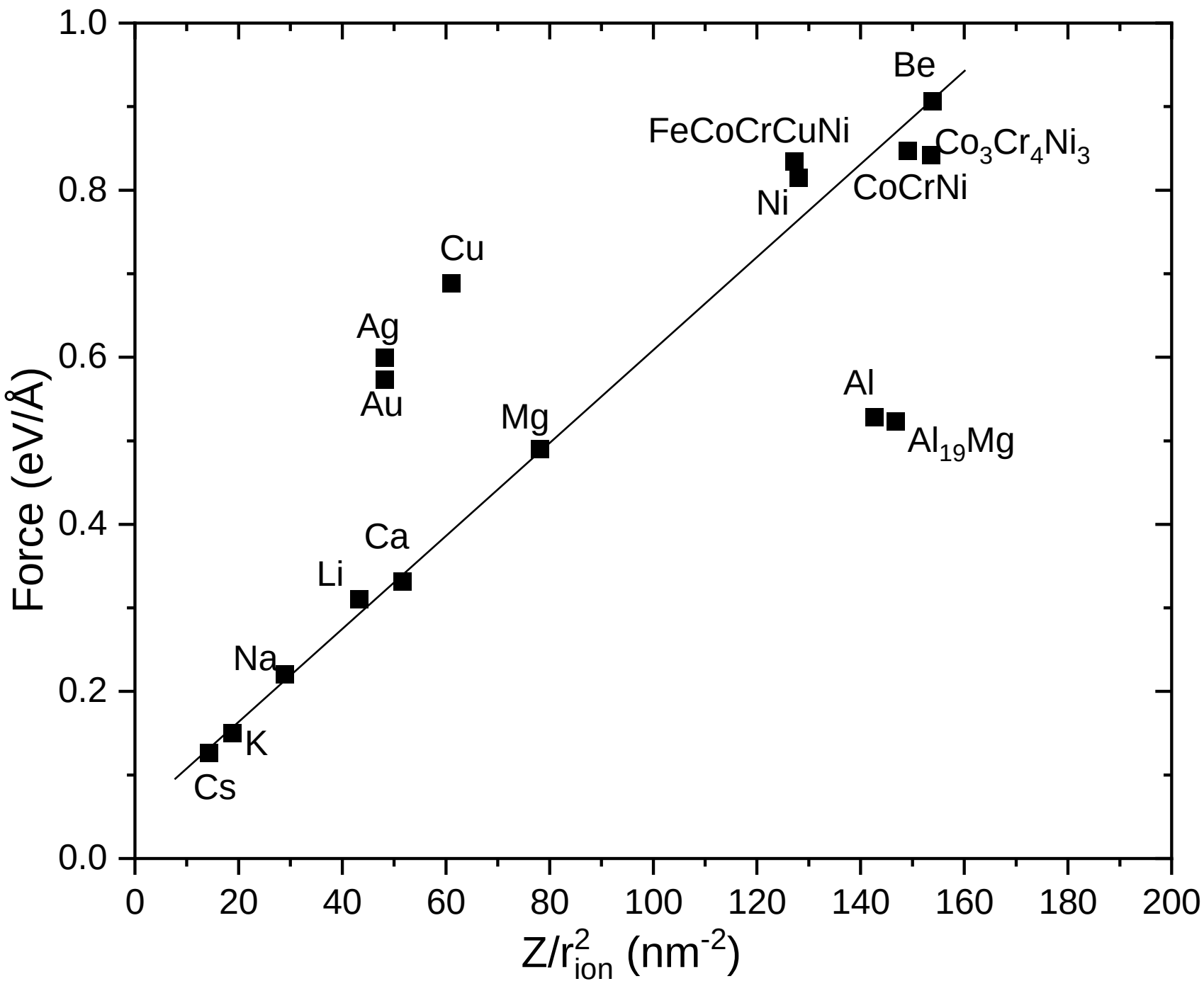

Fig.12. Comparison of the average force per atom with the electrostatic parameter $Z/r^2_{ion}$.

Force is a measure of bonding strength and reflects the atomic bonding capability. The average force computed via LAMMPS provides a qualitative description of interatomic interactions, rather than an exact pairwise atomic force. It should be emphasized that a larger force indicates stronger bonding only when comparing different materials. For a given material, a larger force corresponds to a smaller slope, as shown in Fig. 8 and Fig. 9. That may partly explain why Young's modulus decreases at elevated temperatures.

4. Conclusion:

In this study, molecular dynamics simulations were employed to systematically investigate the influence of surface effect, crystallographic orientation and temperature on the tensile deformation mechanisms and mechanical properties of FCC CoCrNi nanowires.

To characterize the atomic scale deformation mechanism of nanowires, the spatial distributions of potential energy and stress were analyzed to elucidate the surface effects. Surface atoms possess higher potential energy and bear greater stress compared to internal atoms, which enables dislocation nucleation and emission preferentially at the free surfaces of CoCrNi nanowires.

CoCrNi nanowires exhibit significant anisotropy along the [100], [110], and [111] crystallographic orientations. The Young's modulus follows the sequence of $E_{[111]} > E_{[110]} > E_{[100]}$, which is related to the atomic stacking configuration along the loading direction. While the yield strength follows the order: $\sigma_{y[111]} > \sigma_{y[100]} > \sigma_{y[110]}$. This anisotropic yield behaviour depends on the deformation mechanism, Schmid factor, and the normal stress effect.

High temperatures thermally activate the dislocation nucleation process, thereby reducing the yield strength of [100]-oriented CoCrNi nanowires. The Young's modulus exhibits a slight decrease as temperature rises. As shown in Fig.9, the slope of the force declines with increasing temperature, which may account for the temperature induced decrease of Young's modulus.

This work establishes a correlation between the macroscopic mechanical properties and the average atomic force from a microscopic perspective. The average force per atom serves as a qualitative estimation of interatomic bonding strength, which fundamentally determines the Young's modulus of various materials. It is also a key factor influencing yield strength, while other factors such as deformation mechanism and stacking fault energy exert additional influences. Given the strong correlation between macroscopic mechanical parameters and atomic force, the macroscopic properties of materials can be tailored by regulating the average force per atom.
Furthermore, the factors influencing metallic bonding strength were investigated in this study. The electrostatic factor is a minor factor compared to the electronic structure factor. Understanding the relationship between electronic structure and metallic bonding strength-whether between like or dissimilar atoms-is essential for the design of alloy nanowires.

**Acknowledgements**
The authors acknowledge the financial support from the National Natural Science Foundation of China (Grant No. 11704194). The computer time at the National Supercomputer Center in Guangzhou (NSCC-GZ) is gratefully acknowledged.